# On chip, high-sensitivity thermal sensor based on high-Q polydimethylsiloxane-coated microresonator

Bei-Bei Li, Qing-Yan Wang, Yun-Feng Xiao[*], Xue-Feng Jiang, Yan Li, Lixin Xiao and Qihuang Gong[#]

State Key Laboratory of Mesoscopic Physics, School of Physics, Peking University, Beijing 100871, P. R. China

**ABSTRACT**

A high-sensitivity thermal sensing is demonstrated by coating a layer of polydimethylsiloxane (PDMS) on the surface of a silica toroidal microresonator on a silicon wafer. Possessing high-Q whispering gallery modes (WGMs), the PDMS-coated microresonator is highly sensitive to the temperature change of the surroundings. We find that, when the PDMS layer becomes thicker, the WGM experiences a transition from red- to blue-shift with temperature increasing due to the negative thermal-optic coefficient of PDMS. The measured sensitivity (0.151 nm/K) is one order of magnitude higher than pure silica microcavity sensors. The ultra-high resolution of the thermal sensor is also analyzed to reach $10^{-4}$ K.

---

[*] Email address: yfxiao@pku.edu.cn

[#] Email address: qhgong@pku.edu.cn

Whispering-gallery-mode (WGM) microresonators with high-$Q$ factors and small mode volumes have extensive applications in low-threshold lasing [错误!未找到引用源。], nonlinear optics [错误！未找到引用源。], cavity electrodynamics [3-4], cavity optomechanics [5] and miniature sensing [6-8]. In sensor application, for example, when biomolecules are binding on the cavity surface, the resonance shift is typically regarded as the sensing signal. Unfortunately, thermal fluctuations caused by either the ambient temperature variation or probe light field are inevitably involved in the measurement, though some schemes that can suppress the thermal-optic noise have been proposed [9-11]. On the other hand, however, the ultrasensitive change of the WGM resonance to the surrounding temperature can be utilized to design a high-sensitivity thermal sensor.

The temperature-induced resonance shifts in different optical microresonators have been realized recently [12-14]. In Refs. [12-13], the thermal sensing is demonstrated based on silicon or silica microring. The relatively high sensitivity has been obtained but the temperature resolution keeps low because their $Q$ factors are typically lower than $10^4$. A high-$Q$ PDMS microsphere is also demonstrated for high-sensitivity thermal sensing [14]. However, microspheres are difficult to be integrated on chips, which is a great disadvantage when applied in integrated optics. In this paper, we experimentally demonstrate an on-chip, high-sensitivity thermal sensor based on a high-$Q$ polydimethylsiloxane (PDMS)-coated microresonator. PDMS is chosen for its low attenuation loss, good chemical stability and large thermal nonlinearity, which provides the high sensitivity of sensing. High-$Q$ WGMs in the microresonator predict a high resolution of the resonance shift (thus a low detection limit).

To fabricate such a PDMS-coated microresonator, a silica microtoroid with major (minor) diameter of 36 (7) μm and is nanometer-sized moved to touch PDMS droplets (RTV615 A and B,

5:1) prepared on a fiber taper. Due to low surface tension, the liquid PDMS mainly spreads out along the peripheral ring surface of the microtoroid and form a smooth thin layer on the toroid surface [11]. Finally, it takes 4-6 hours for the PDMS coating to cure at a certain temperature. The coating thickness is generally controlled by the size of PDMS droplet. Compared with the spin coating technique [15], the present method allows for selectively coating a single microcavity. In our experiment, the increased thickness of the PDMS layer for each coating is about 100 nm. Figures 1(a) and 1(b) show optical images of the toroid before and after several coating processes, respectively. It is obvious that the minor diameter of the toroid increases. In addition, the roughness of PDMS surface with this coating process is nanometer sized, as demonstrated in Fig. 1(c), where an atomic force microscopy (AFM) image of a portion of PDMS surface is shown.

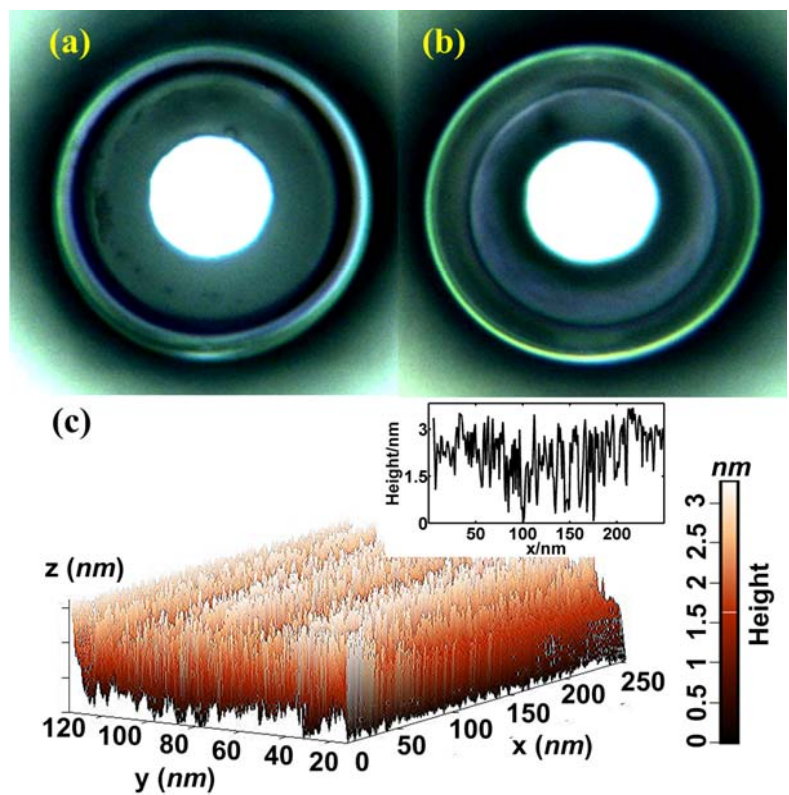

FIG. 1. (Color online) Top views of a silica microtoroid before (a) and after (b) PDMS coating. An obvious increase of the minor diameter can be observed. (c) AFM image of a portion of peripheral ring surface of the coated microtoroid. Inset: The section profile along the $x$-direction, showing a roughness of ~3 nm.

To study the resonant wavelength shifts of the PDMS-coated silica microtoroid, a fiber taper with diameter ~ 1.5 μm is used to couple the laser light (in 1550 nm band) into the PDMS-coated toroidal microresonator and also collect the light out. To minimize the thermal effect induced by the input light itself, the input power is kept below 10 μW. The transmission spectrum of the fiber taper is detected by a photoreceiver, and finally monitored by an oscilloscope. A thermalelectric cooler is applied to change the surrounding temperature of the microresonator. At different coating thickness, we measure the resonance shifts of the fundamental WGM with the surrounding temperature changing from room temperature to several degrees higher.

The inset of Fig. 2 shows the transmission spectra of a fundamental WGM at several surrounding temperatures with the coating thickness of about 1.6 μm. Under each temperature, we maintain the same taper-microresonator coupling efficiency in order that the thermal effects caused by the laser power are the same for each time. Figure 2 shows the resonant wavelength shifts of the fundamental mode versus temperature change with various coating thicknesses. Excellent linear dependence of the resonant wavelength shift against the temperature change is observed. The result further shows that with the increasing coating thickness, the fundamental WGM experiences a transition from red- to blue-shift when the temperature increases. The largest sensitivity of the fundamental WGM to temperature change is 0.151 nm/K, which is about 12 times of that based on the pure silica toroidal microresonator. In addition, with the coating thickness increasing, we find that the $Q$ factor slightly decreases from $8.5 \times 10^6$ to $1.5 \times 10^6$ because of the absorption loss of PDMS.

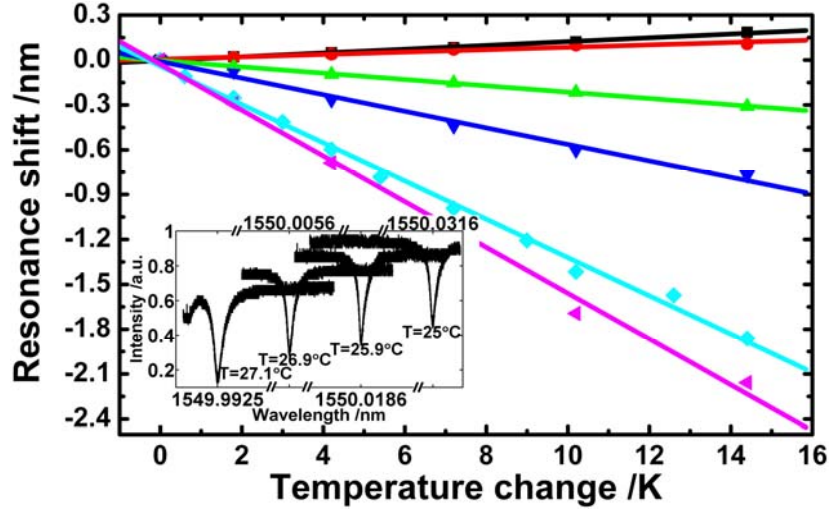

FIG. 2. (Color online) Resonant wavelength shift vs. temperature change based on room temperature at several PDMS coating thicknesses. From the top to bottom, the thickness gradually increases. The solid lines are linear fittings. Inset: The transmission spectra of a fundamental WGM for several temperatures when the coating thickness is about 1.6 μm

To analyze the experimental results, the thermal effect of this PDMS coated toroidal resonator is theoretically investigated. The resonant wavelength change of a fundamental WGM versus temperature can be given by

$$\frac{d\lambda}{dT} = \lambda_r \left( \frac{1}{n_{\text{eff}}} \frac{dn_{\text{eff}}}{dT} + \frac{1}{D} \frac{dD}{dT} \right) \tag{1}$$

where, $\lambda_r$ is the cold cavity resonant wavelength, $n_{\text{eff}}$ and $D$ represent the effective refractive index (RI) and the principal diameter of the coated microtoroid. In this hybrid system, the effective RI can be approximately expressed as $n_{\text{eff}} = \eta_1 n_{\text{silica}} + \eta_2 n_{\text{PDMS}} + \eta_3 n_{\text{air}}$, where $\eta_1$, $\eta_2$, and $\eta_3$ represent the energy fractions of the WGM field in silica, PDMS and air, respectively. The diameter $D$ consists of the principal diameter of the silica microtoroid $d$ and the thickness of the

PDMS film $t$, expressed by $D=d+2t$.

Before studying the resonant shift of WGM dependent on the coating thickness, we note that the observed highest blue-shift rate is one order of magnitude higher than that based on pure silica microcavity sensor. This is unreachable based on the parameters in Ref. [11]. The most potential reason is that the curing temperature and the ratio of RTV615 A and B result in change of the thermal-optic coefficient $dn_{PDMS}/dT$ of PDMS. To accord with this large blue-shift rate, here $dn_{PDMS}/dT$ is chosen as $-1.8×10^{-4}$ while the thermal-expansion coefficient is $2.7×10^{-4}$. These two values match well with both the highest blue-shift rate (shown in this paper) and the highest red-shift rate (in the case of a pure PDMS microcavity, not shown here).

By using a finite element method (FEM), we can calculate the field distributions of WGMs, i.e., to obtain $\eta_{1,2,3}$. The solid curve in Fig. 3 plots the energy fraction $\eta_2$ in PDMS which gradually increases with the thicker coating. For example, when the coating thickness $t = 2.25$ μm, $\eta_2$ is already 0.97. According to Eq. (1), the sensitivity of the wavelength shift versus temperature change $d\lambda/dT$ can be obtained, as shown in the dashed curve. When the thickness is small, the first term in Eq. (1) is dominant and the positive thermal refraction of silica is partly compensated by the negative thermal refraction of PDMS, leading to the decrease of the red-shift rate. With further increasing of PDMS thickness, the WGM experiences a transition from red- to blue-shift (i.e., over compensated), and the blue-shift rate is much larger than the red-shift. When the thickness increases furthermore (above 2.25 μm), the second term in Eq. (1) related to thermal expansion effect becomes dominant, and thus decreases the blue-shift. With FEM, we can also directly obtain the resonant wavelength for different environment temperature, as shown in the dotted curve. This simulated result is close to the analytical result obtained by Eq. (1). The slight deviation is attributed to the inaccuracy of the effective diameter in Eq. (1) from the actual one. Both the experimental and theoretical results show that the PDMS-coated silica microcavity can provide a highly sensitive detection of the environment temperature (up to 0.151nm/K) at a proper coating thickness (~2.25 μm).

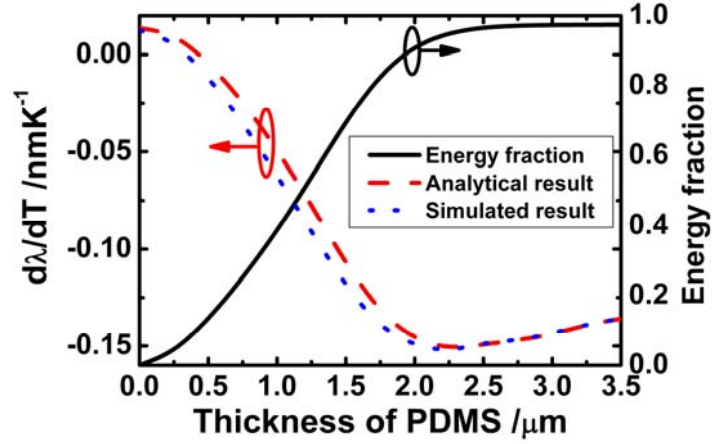

FIG. 3. (Color online) The energy fraction in PDMS vs. different PDMS thicknesses (black solid curve). The sensitivity obtained from Eq. (1) (red dashed curve) and the numerical simulation (blue dotted curve). Here, the thermal-optic and expansion coefficients for PDMS (silica) are $-1.8 \times 10^{-4}$ and $2.7 \times 10^{-4}$ ($5.5 \times 10^{-7}$ and $1.19 \times 10^{-5}$), respectively. The refraction indexes of PDMS and silica are 1.41 and 1.45.

We now turn to discuss the detection limit based on this PDMS-coated microresonator because not only the detection sensitivity but also the resolution represents the important performance of the sensing. Suppose the instrument can resolve the wavelength at a resolution of $\Delta \lambda_{min}$, the smallest measurable temperature change required to create $\Delta \lambda_{min}$ is then obtained as

$$(\Delta T)_{min} = \frac{(\Delta \lambda)_{min}}{d\lambda / dT} = \frac{(\Delta \lambda)_{min} \lambda_r}{\Delta \lambda_r Q d\lambda / dT} \quad (1)$$

where $\Delta \lambda_r$ is the linewidth of the resonance. As the sensor detection limit is set by how well one can locate $\lambda_r$, a high $Q$ is essential. The smallest detectable wavelength shift $(\Delta \lambda)_{min} / \Delta \lambda_r$ is limited

by the detector noise, and can be expressed as (see Appendix)

$$(d\lambda)_{\min} = \frac{[S_{\text{Thermal}}(f) + S_{\text{Shot}}(f)]^{1/2}}{[\dfrac{16\eta R^{3/2} P_0 Q^4 B^2}{Q_c Q_0 \lambda_r (f_r^2 + 4Q^2 B^2)}]} \tag{3}$$

where $S_{\text{Thermal}}$ and $S_{\text{Shot}}$ denote the thermal noise and the shot noise of the detector, respectively. $P_0$ is the input power. $R$, $\eta$ and $B$ are the load resistance, the photoelectric conversion efficiency and the bandwidth of the photoreceiver, respectively. $Q_0$ and $Q_c$ are the intrinsic quality factor and loaded quality factor. The detection limit of temperature of the sensor can be obtained according to Eqs. (2) and (3). The influence of the detector noises on the detection limit with $Q=1\times10^6$ is plotted in Fig. 4. With $P_0<0.5$ μW the smallest detectable temperature change is mainly limited by thermal noise, while with $P_0$ increasing, the shot noise becomes dominant. The detection in our experiment occurs in the shot-noise-limited regime, which allows determining a wavelength shift with a precision of almost two orders of magnitude higher than given by the linewidth [16]. Thus the detection limit of temperature can be obtained as $(\Delta T)_{\min} = 1\times10^{-4}$K, which is nearly half of the reported resolution of the PDMS microsphere sensor [14].

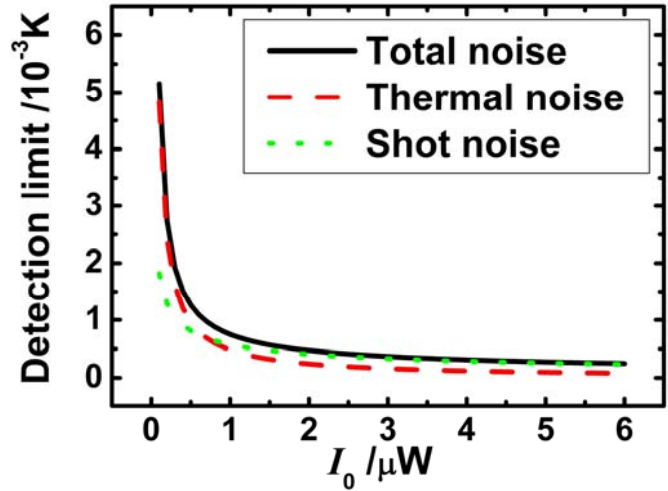

FIG. 4. (Color online) Detection limit of the PDMS-coated microresonator thermal sensor vs. input power.

In summary, we propose an on-chip thermal sensor based on a PDMS-coated silica microtoroid which can be fabricated using a simple but highly efficient method. The thermal responses of the fundamental WGMs are experimentally and theoretically investigated under different coating thickness. The results show that such a hybrid microresonator is a promising thermal sensor with high sensitivity (0.151 nm/K) and high resolution ($1 \times 10^{-4}$K). Moreover, its on-chip feature is desirable to fulfill the demand for integration and miniaturization in optics.

BBL and QYW contributed equally in this work. The authors acknowledge financial support from the National Natural Science Foundation of China under Grant No. 10821062, the National Basic Research Program of China under Grant Nos. 2006CB921601, 2007CB307001 and 2009CB930504. YFX was also supported by the Research Fund for the Doctoral Program of Higher Education (No.20090001120004) and the Scientific Research Foundation for the Returned Overseas Chinese Scholars.

**APPENDIX**

Consider a cavity-taper coupling system. The output fields of the cavity can be expressed as

$$a_o = \frac{[i(\omega - \omega_r) - \kappa/2 + \kappa_c]a_i}{i(\omega - \omega_r) - \kappa/2}. \tag{A1}$$

where $a_i$ and $a_o$ are the amplitudes of the input and onput fields; $\kappa = \kappa_0 + \kappa_c$ describes the total dissipation of the mode in the cavity. $\kappa_0 = \omega_0/Q_0$ and $\kappa_c = \omega_0/Q_c$ are the dissipation coefficients of the mode due to the intrinsic loss of the cavity and the insertion loss of the taper, respectively. Then the received light intensity of the photoreceiver is

$$I = \int_{-\infty}^{+\infty} |a_o(f)|^2 df \cong P_0[B - \frac{2f_r Q}{Q_c Q_0}\arctan(\frac{2QB}{f_r})]. \tag{A2}$$

where $B$ is the bandwidth of the photoreceiver. When the resonance frequency of the cavity $f_r$ changed by $df_r$, the received intensity of photoreceiver changes by

$$\frac{dI}{df_r} = \frac{16 P_0 Q^4 B^3}{Q_c Q_0} \frac{1}{f_r(f_r^2 + 4Q^2 B^2)} \tag{A3}$$

Thus the change of the photoreceiver output signal produced by the change of the light intensity is

$$V = \eta R dI/B = \frac{16\eta R P_0 Q^4 B^2}{Q_c Q_0 \lambda_r} \frac{d\lambda}{(f_r^2 + 4Q^2 B^2)} \tag{A4}$$

The signal to noise ratio (SNR) of the thermal sensor can be expressed as

$$\text{SNR} = \frac{V^2 R}{S_{\text{Thermal}}(f) + S_{\text{Shot}}(f)} \tag{A5}$$

When SNR=1, we can deduce the smallest detectable wavelength shift under the limitation of noise, and write it as

$$(d\lambda)_{\min} = \frac{\left[S_{\text{Thermal}}(f) + S_{\text{Shot}}(f)\right]^{1/2}}{\left[\dfrac{16\eta R^{3/2} P_0 Q^4 B^2}{Q_c Q_0 \lambda_r (f_r^2 + 4Q^2 B^2)}\right]} \qquad (\text{A6})$$